\documentstyle[11pt,amssymb]{article}

\def\dalemb#1#2{{\vbox{\hrule height .#2pt
        \hbox{\vrule width.#2pt height#1pt \kern#1pt
                \vrule width.#2pt}
        \hrule height.#2pt}}}

\def\0{{\sst{(0)}}}
\def\1{{\sst{(1)}}}
\def\2{{\sst{(2)}}}
\def\3{{\sst{(3)}}}
\def\4{{\sst{(4)}}}
\def\5{{\sst{(5)}}}
\def\6{{\sst{(6)}}}
\def\7{{\sst{(7)}}}
\def\8{{\sst{(8)}}}

\def\ep{\epsilon}

\let\a=\alpha \let\b=\beta

\def\nn{\nonumber} \def\bd{\begin{document}} \def\ed{\end{document}}
\def\ds{\documentstyle} \let\fr=\frac \let\bl=\bigl \let\br=\bigr
\let\Br=\Bigr \let\Bl=\Bigl 
\let\bm=\bibitem
\let\na=\nabla
\let\pa=\partial \let\ov=\overline 
\newcommand{\be}{\begin{equation}} 
\newcommand{\ee}{\end{equation}} 
\def\ba{\begin{array}}
\def\ea{\end{array}}
\def\ft#1#2{{\textstyle{{\scriptstyle #1}\over {\scriptstyle #2}}}}
\def\fft#1#2{{#1 \over #2}}
\def\del{\partial}
\def\sst#1{{\scriptscriptstyle #1}}
\def\oneone{\rlap 1\mkern4mu{\rm l}}
\def\ie{{\it i.e.\ }}
\def\via{{\it via}}
\def\semi{{\ltimes}}
\def\str{{\rm str}}
\def\jm{{\rm j}}
\def\im{{\rm i}}
\def\mapright#1{\smash{\mathop{-\!\!\!-\!\!\!-\!\!\!-\!\!\!-\!\!\!
             \longrightarrow}\limits^{#1}}}
\def\maprightt#1#2{\smash{\mathop{-\!\!\!-\!\!\!-\!\!\!-\!\!\!-\!\!\!
             \longrightarrow}\limits^{#1}_{#2}}}

\newcommand{\ho}[1]{$\, ^{#1}$}
\newcommand{\hoch}[1]{$\, ^{#1}$}
\newcommand{\bea}{\begin{eqnarray}} 
\newcommand{\eea}{\end{eqnarray}} 
\newcommand{\ra}{\rightarrow}
\newcommand{\lra}{\longrightarrow}
\newcommand{\Lra}{\Leftrightarrow}
\newcommand{\ap}{\alpha^\prime}
\newcommand{\bp}{\tilde \beta^\prime}
\newcommand{\tr}{{\rm tr} }
\newcommand{\Tr}{{\rm Tr} } 
\newcommand{\NP}{Nucl. Phys. }
\newcommand{\tamphys}{\it Center for Theoretical Physics\\
Texas A\&M University, College Station, Texas 77843}
\newcommand{\ens}{\it Laboratoire de Physique Th\'eorique de l'\'Ecole
Normale Sup\'erieure\hoch{2,3}\\
24 Rue Lhomond - 75231 Paris CEDEX 05}
\newcommand{\upenn}{\it Department of Physics and Astronomy\\
University of Pennsylvania, Philadelphia, Pennsylvania 19104}

\newcommand{\auth}{M. Cveti\v{c}\hoch{\dagger\star1}, 
H. L\"u\hoch{\dagger1} and and C.N. Pope\hoch{\ddagger\star2}}

\begin{document}
\begin{flushright}
\hfill{CTP TAMU-27/99}\\
\hfill{UPR-851-T}\\
\hfill{hep-th/9906221}\\
\hfill{June 1999}\\
\end{flushright}


\begin{center}
{ \large {\bf Gauged Six-dimensional Supergravity from Massive Type IIA 
}}

\vspace{15pt}
\auth

\vspace{15pt}

{\hoch{\dagger}\upenn}

\vspace{15pt}
{\hoch{\ddagger}\tamphys}

\vspace{15pt}
{\hoch{\star}
\it SISSA, Via Beirut No. 2-4, 34013 Trieste, Italy}

\vspace{40pt}

\underline{ABSTRACT}
\end{center}

     We obtain the complete non-linear Kaluza-Klein ansatz for the
reduction of the bosonic sector of massive type IIA supergravity to
the Romans $F(4)$ gauged supergravity in six dimensions.  The
latter arises as a consistent warped $S^4$ reduction.

{\vfill\leftline{}\vfill
\footnoterule
{\footnotesize \hoch{1} Research supported in part by DOE grant 
DE-FG02-95ER40893 \vskip -12pt} \vskip 14pt
{\footnotesize  \hoch{2} Research supported in part by DOE 
grant DE-FG03-95ER40917.\vskip  -12pt}}

\pagebreak
\setcounter{page}{1}

    The conjectured duality between supergravity on an anti-de Sitter
(AdS) background and a superconformal field theory (CFT) on its boundary
\cite{mald,gkp,wit2,mald2} has led to a renewed interest in the mechanism
whereby the relevant gauged supergravities can be obtained by Kaluza-Klein 
reduction from higher dimensions.  It has long been known that the
maximal gauged theories in $D=4$ and $D=7$ can be obtained by
reduction of eleven-dimensional supergravity on $S^7$ or $S^4$, and
that the maximal gauged theory in $D=5$ can be obtained from an $S^5$
reduction from type IIB supergravity.  In each case, it is believed
that the reduction is consistent, in the sense that the reduction
ansatz, with its truncation to the fields of the supergravity
multiplet, satisfies the higher-dimensional equations of motion
provided that the lower-dimensional equations of motion are
satisfied.  This is important in the context of the AdS/CFT
correspondence, since it implies that massive fields can be ignored
when calculating correlation functions in the conformal field theory
\cite{wit2}.   

    A long-standing puzzle has been to obtain a higher-dimensional
Kaluza-Klein interpretation for the gauged supergravity theory in six
dimensions \cite{romans6}, whose dual description on its boundary is
a five-dimensional $N=2$ superconformal field theory
\cite{seiberg,intrill}.  It was suggested in \cite{ferrara} that it
could be related to the ten-dimensional massive type IIA theory
\cite{romanstype2a}.  Recently, it was shown that the massive IIA
theory admitted an AdS$_6\times S^4$ solution, with a
``warped-product'' metric \cite{oz}.  This was derived as the
near-horizon limit of a localised D4-D8 brane configuration
\cite{youm}.

    In this letter we resolve the puzzle, by obtaining the complete
non-linear bosonic Kaluza-Klein ansatz for the reduction of the
massive IIA theory on $S^4$, and showing that it gives a consistent
truncation to the six-dimensional gauged theory.  Our starting point
is the bosonic Lagrangian of massive type IIA supergravity
\cite{romanstype2a}.  In the language of differential forms, it is
given by \cite{branetension}
\bea
{\cal L}_{10} &=& \hat R\, {\hat *\oneone} -
\ft12 {\hat *d\hat \phi}\wedge d\hat \phi
- \ft12 e^{\fft32\hat\phi}\, {\hat *\hat F_\2}\wedge \hat F_\2 - 
\ft12 e^{-\hat \phi}\, {\hat *\hat F_\3}\wedge \hat F_\3- 
\ft12 e^{\fft12\hat\phi}\, {\hat *\hat F_\4}\wedge \hat F_\4 \nn\\
&&\!\!
-\ft12 d\hat A_\3\wedge d\hat A_\3 \wedge \hat A_\2 - \ft16 m\, 
d\hat A_\3 \wedge (\hat A_\2)^3
 -\ft1{40} m^2\, (\hat A_\2)^5 -\ft12 m^2\, e^{\fft52\hat \phi}\, 
{\hat *\oneone}\,,\label{romans1}
\eea
where the field strengths are given in terms of potentials by
\bea
\hat F_\2 &=& d\hat A_\1 + m\, \hat A_\2\ ,\qquad \hat F_\3 = 
d\hat A_\2\,,\nn\\
\hat F_\4 &=& d\hat A_\3 + \hat A_\1\wedge d\hat A_\2 + \ft12 m\, 
\hat A_\2\wedge \hat A_\2\,.\label{romfields}
\eea
In the above, we have used the hat symbol to denote the
ten-dimensional fields and Hodge dual, and subscripts on form fields
indicate the degrees of the forms.  It follows that the equations
of motion and Bianchi identities are
\bea
&&d(e^{\fft12\hat \phi}\, {\hat *\hat F_\4}) = -\hat F_\3\wedge \hat
F_\4\ ,\qquad d(e^{\fft32\hat \phi}\, {\hat * \hat F_\2}) =-
e^{\fft12\hat \phi}\, {\hat *\hat F_\4}\wedge \hat F_\3\,,\nn\\
&&d(e^{-\hat\phi}\, {\hat * \hat F_\3}) = -\ft12 \hat F_\4\wedge \hat
F_\4 -m\, e^{\fft32\hat\phi}\, {\hat * \hat F_\2}-
e^{\fft12\hat \phi}\, {\hat * \hat F_\4}\wedge \hat F_\2\,,\nn\\
&&d{\hat * d\hat \phi}= -\ft54m^2\, e^{\fft52\hat\phi}\, 
{\hat *\oneone}-\ft34 e^{\fft32\hat \phi}\, {\hat *\hat F_\2}\wedge
F_\2 + \ft12 e^{-\hat \phi}\, {\hat *\hat F_\3}\wedge \hat F_\3
-\ft14 e^{\fft12\hat\phi}\, {\hat *\hat F_\4}\wedge\hat F_\4\,,\nn\\
&&d\hat F_\4=\hat F_\2\wedge \hat F_\3\,,\qquad
d\hat F_\3=0\,,\qquad d\hat F_\2=m\,\hat F_\3\,,
\label{d10eom}
\eea
for the form fields and dilaton, together with the Einstein equation
(in vielbein components)
\bea
\hat R_{AB} &=& \ft12 \del_A\hat\phi\, \del_B\hat\phi + 
\ft1{16}m^2\, e^{\fft52\hat\phi}\, \eta_{AB} + \ft1{12}
e^{\fft12\hat\phi}\, \Big( \hat F^2_{\4 AB} -\ft3{32} \hat F_\4^2\,
\eta_{AB} \Big)\label{d10einst}\\
&& + \ft14 e^{-\hat\phi}\, \Big( \hat F^2_{\3 AB} -\ft1{12} \hat
F_\3^2\, \eta_{AB}\Big) + \ft12 e^{\fft32\hat\phi}\, \Big( \hat
F^2_{\2 AB} -\ft1{16} \hat F_\2^2\, \eta_{AB}\Big)\,.\nn
\eea

    We shall now describe how we can perform a 4-sphere reduction of
the massive IIA theory, with a consistent truncation to the fields
of gauged $N=1$, $D=6$ supergravity.  The bosonic fields in
this theory comprise the metric, a dilaton, a 2-form potential and a
1-form potential, together with the gauge potentials of $SU(2)$
Yang-Mills.  The bosonic Lagrangian \cite{romans6}, converted to
the language of differential forms, is
\bea
{\cal L}_6 &=& R\, {*\oneone} -\ft12 {*d\phi}\wedge d\phi
- g^2\Big(\ft29 e^{\fft{3}{\sqrt2}\phi} -\ft83 e^{\fft1{\sqrt2}\phi} -
2 e^{-\fft1{\sqrt2}\phi}\Big)\,  {*\oneone}\nn\\
&&-\ft12 e^{-\sqrt2\phi}\, {*F_\3\wedge F_\3} -\ft12
e^{\fft1{\sqrt2}\phi}\, \Big( {*F_\2}\wedge F_\2 + {*F_\2^i}\wedge
F_\2^i \Big) \label{d6lag}\\
&& - A_\2\wedge(\ft12
dA_\1\wedge dA_\1 +\ft13 g\, A_\2\wedge dA_\1 +\ft2{27}  
g^2\, A_\2\wedge A_\2 +\ft12 F_\2^i\wedge F_\2^i)\,,\nn
\eea
where $F_\3=dA_\2$, $F_\2= dA_\1 + \ft23g\, A_\2$ and $F_\2^i =
dA_\1^i + \ft12 g\, \ep_{ijk} A_\1^j\wedge A_\1^k$.  Here $*$ is the
six-dimensional Hodge dual. 

    We find that the reduction ans\"atze for the metric, form fields
and dilaton of the ten-dimensional massive type IIA theory are
\bea
d\hat s_{10}^2\!\! &=&\!\! (\sin\xi)^{\fft1{12}}\, X^{\fft18}\Big[
\Delta^{\fft38}\, ds_6^2 + 2g^{-2}\, \Delta^{\fft38}\, X^2\, d\xi^2
+\ft12g^{-2}\, \Delta^{-\fft58}\, X^{-1}\, \cos^2\xi
\sum_{i=1}^3(\sigma^i - g\, A_\1^i)^2\Big]\,,\nn\\
\hat F_\4 &=& -\ft{\sqrt2}{6}\, g^{-3}\, s^{1/3}\, c^3\, \Delta^{-2}\,
U\, d\xi\wedge\ep_\3 -\sqrt2 g^{-3}\, s^{4/3}\, c^4\, \Delta^{-2}\,
X^{-3}\, dX\wedge \ep_\3 \nn\\
&&-\sqrt2 g^{-1}\, 
s^{1/3}\, c\, X^4\, {*F_\3}\wedge d\xi
-\ft1{\sqrt2} s^{4/3}\, X^{-2}\, {*F_\2} \nn\\
&& +\ft1{\sqrt2} g^{-2}\,
s^{1/3}\, c\, F_\2^i \, h^i\wedge d\xi -\ft1{4\sqrt2} g^{-2}\,
s^{4/3}\, c^2\, \Delta^{-1}\, X^{-3}\,  F_\2^i \wedge
h^j\wedge h^k\, \ep_{ijk}\,,\label{fans}\\
\hat F_\3 &=& s^{2/3}\, F_\3 + g^{-1}\, s^{-1/3}\, c\, F_\2\wedge d\xi
\,,\nn\\
\hat F_\2 &=& \ft1{\sqrt2}\, s^{2/3}\, F_\2\,,\qquad
e^{\hat\phi} = s^{-5/6}\, \Delta^{1/4}\, X^{-5/4}\,,\nn
\eea
where $X$ is related to the dilaton $\phi$ in (\ref{d6lag}) by
$X=e^{-\fft1{2\sqrt2}\phi}$, and
\bea
\Delta &\equiv & X\cos^2\xi +X^{-3} \sin^2 \xi\,,\nn\\
U &\equiv& X^{-6}\, s^2 - 3 X^2\, c^2 + 4 X^{-2}\, c^2 - 6 X^{-2}\,.
\eea
The quantities $\sigma^i$ are left-invariant 1-forms on $S^3$, which
satisfy $d\sigma^i = -\ft12 \ep_{ijk}\, \sigma^j\wedge \sigma^k$.  We
have also defined $h^i\equiv \sigma^i-g\, A_\1^i$, $\ep_\3\equiv
h^1\wedge h^2\wedge h^3$, and $s=\sin\xi$ and $c=\cos\xi$.  The gauge
coupling constant $g$ is related to the mass parameter $m$ of the
massive type IIA theory by $m= \ft{\sqrt2}{3}\, g$.

   It is useful also to present the expressions for the
ten-dimensional Hodge duals of the form fields given above, and for
$d\hat\phi$.  We find that they are given by
\bea
e^{\fft12\hat\phi}\, {\hat *\hat F_\4} &=& -\ft{\sqrt2}{3} g\, U\,
\ep_\6 + 4 \sqrt2 g^{-1}\, s\, c\, X^{-1}\, {*dX}\wedge d\xi\nn\\
&&-\ft{\sqrt2}{4} g^{-3}\, c^4\, \Delta^{-1}\, X\, F_\3\wedge \ep_\3 
 +\ft1{2\sqrt2} g^{-4}\, s\, c^3\, \Delta^{-1}\, X^{-3}\, F_\2\wedge
d\xi \wedge \ep_\3\nn\\
&&-\ft1{4\sqrt2} g^{-2}\, c^2\, X^{-2}\,{*F_\2^i}\wedge h^j\wedge h^k
\, \ep_{ijk} + \ft1{\sqrt2} g^{-2}\, s\, c\, X^{-2}\, 
{*F_\2^i}\wedge h^i\wedge
d\xi\,,\nn\\
e^{-\hat\phi}\, {\hat *\hat F_\3} &=& \ft12 g^{-4}\, s^{5/3}\, c^3\,
\Delta^{-1}\, X\, {* F_\3}\wedge d\xi\wedge \ep_\3  
-\ft14 g^{-3}\, s^{2/3}\, c^4\, \Delta^{-1}\, X^{-1}\, {*F_\2} \wedge
\ep_\3 \,,\nn\\
e^{\fft32\hat\phi}\, {\hat *\hat F_\2} &=& \ft1{2\sqrt2} g^{-4}\,
s^{-1/3}\, c^3\, X^{-2}\, {*F_\2}\wedge d\xi\wedge 
\ep_\3\,,\label{hodgedual}\\
{\hat *d\hat\phi} &=& -\ft12 g^{-4}\, s^{1/3}\, c^3\, (X\, c^2 + 2
X^{-3}\, s^2) \, \Delta^{-1}\, X^{-1}\, {*dX}\wedge d\xi\wedge \ep_\3 
\nn\\
&&+\ft1{16} g^{-2}\, s^{1/3}\, c^3\, (\Delta^{-1}\, \del_{\xi}\Delta
-\ft{10}3 \cot\xi)X^{-2}\, \ep_\6\wedge \ep_\3\,,\nn
\eea
where $\ep_\6$ is the volume form of the metric $ds_6^2$.

   If we set $X=1$ and $A_\1^i=0$, then
$ds_6^2$ becomes an Einstein metric, which can, for example, be
AdS$_6$.  In this case, the ten-dimensional geometry becomes
AdS$_6\times S^4$ with a warp factor \cite{oz},
\be
ds_{10}^2=(\sin\xi)^{\fft1{12}}\, \Big[ ds^2_{\rm{AdS}_6}  +
2g^{-2}\, \Big(d\xi^2 + \ft14\cos^2\xi\, \sum_i (\sigma^i)^2 
\Big)\Big]\,,
\ee
which is the near-horizon limit \cite{oz} of a localised D4-D8 brane
configuration \cite{youm}.\footnote{To be more precise, the $S^4$ here
is not really the entire 4-sphere, but rather just the upper
hemisphere of a 4-sphere, viewed as a foliation of 3-spheres
\cite{oz}.  This is because the conformal warp factor
$(\sin\xi)^{1/12}$ approaches zero as the ``latitude'' coordinate
$\xi$ approaches the equatorial 3-sphere at $\xi=0$, thus defining a
boundary to the 4-manifold.}  The configuration is a solution of the
massive type IIA theory, where the AdS$_6$ metric $ds^2_{\rm{AdS}_6}$
has Ricci tensor $R_{ab} = -\ft{10}9 g^2\, g_{ab}$ and the 4-sphere
metric $2g^{-2}\, (d\xi^2 + \ft14\cos^2\xi\, \sum_i(\sigma^i)^2)$ has
Ricci tensor $R_{\a\b} = \ft32 g^2\, g_{\a\b}$.  It follows from the
ansatz (\ref{fans}) that the ten-dimensional fields $\hat F_\4$ and
$\hat\phi$ take the non-vanishing forms
\be
\hat F_\4 = \ft{5\sqrt2}{6}\, g^{-3}\, 
(\sin\xi)^{1/3}\, \cos^3\xi\, d\xi\wedge \ep_\3\,, \qquad
e^{\hat\phi} = (\sin\xi)^{-5/6}\,.
\ee
When the fields $X$ and $A_\1^i$ are excited, $X$ parameterises
inhomogeneous deformations of the 4-sphere, leaving the foliating
3-spheres intact, while $A_\1^i$ describes deformations of the
3-spheres corresponding to right translations under $SU(2)$.

     Substituting the ans\"atze (\ref{fans}) into the equations of
motion and Bianchi identities (\ref{d10eom}) for the form fields and
dilaton of the massive type IIA theory, we find that they are
satisfied provided the six-dimensional fields satisfy the following
equations of motion:
\bea
&& d(X^4\, {* F_\3}) = -\ft12 F_\2\wedge F_\2 - \ft12 F_\2^i\wedge
F_\2^i - \ft23 g\, X^{-2}\, {* F_\2}\,,\nn\\
&& d(X^{-2}\, {*F_\2}) = - F_\2\wedge F_\3\,,\qquad 
D(X^{-2}\, {*F_\2^i}) = - F_\2^i\wedge F_\3\,,\label{d6eom}\\
&& d(X^{-1}\, {*dX}) = \ft18 X^{-2}\, ({*F_\2}\wedge F_\2  + {*F_\2^i}
\wedge F_\2^i) - \ft14 X^4\, {*F_\3}\wedge F_\3 \nn\\
&& \qquad\qquad\qquad\quad
 + g^2\, (\ft16 X^{-6} -\ft23 X^{-2} +\ft12 X^2)\, {*\oneone}\,,\nn
\eea
where $D$ is the Yang-Mills gauge-covariant exterior derivative,
$D\, \omega^i= d\omega^i +g\, \ep_{ijk}\, A_\1^j\wedge \omega^k$.
Note that the Bianchi identities for $\hat F_\3$ and $\hat F_\2$ are
satisfied identically, whilst that for $\hat F_\4$ already implies the
equations of motion for $F_\3$ and $F_\2$.  

          Evaluating the ten-dimensional Einstein equation 
(\ref{d10einst}) with the ans\"atze (\ref{fans}) 
is a more exacting task.  After doing so, we find that consistency
again requires the equations of motion for $F_\2^i$ and $X$ given in
(\ref{d6eom}), and in addition it implies the six-dimensional Einstein
equation
\bea
R_{\mu\nu} &=& 4X^{-2}\, \del_\mu X\, \del_\nu X  
 + g^2\, (\ft1{18} X^{-6} -\ft23 X^{-2} - \ft12 X^2) \, g_{\mu\nu}
+\ft14 X^4\, (F^2_{\3 \mu\nu} - \ft16 F_\3^2\, g_{\mu\nu} )\nn\\
&&+\ft12 X^{-2}\, (F^2_{\2 \mu\nu} - \ft18 F_\2^2\, g_{\mu\nu} )
+ \ft12 X^{-2}\, ( (F_\2^i)^2_{\mu\nu} - \ft18 (F_\2^i)^2) \,
g_{\mu\nu})\,. \label{d6einst} 
\eea
It is now straightforward to see that the full set of 
six-dimensional equations of motion (\ref{d6eom}) and (\ref{d6einst}) 
are precisely those which follow from the Lagrangian (\ref{d6lag}) for
$N=1$, $D=6$ gauged supergravity.   

     In our derivation, the consistency of the reduction ansatz is
definitively established, since we have explicitly substituted it into
the higher-dimensional equations of motion, and shown that these
equations are satisfied if and only if the lower-dimensional equations
of motion are satisfied.  This is, by definition, what one means by a
consistent Kaluza-Klein reduction.  The Kaluza-Klein reduction
procedure is sometimes stated instead at the level of the action;
namely, that one would substitute the ansatz into the
higher-dimensional action, integrate over the internal directions, and
thereby arrive at an action for the lower-dimensional fields.  Of
course in such an approach, it would be necessary to construct an
independent argument for why the reduction ansatz was a consistent
one.  However, there are other reasons also why substituting the
ansatz into the Lagrangian might be problematical.  To illustrate
this, it is instructive to look at the reduction we have considered in
this letter, simplified initially by restricting the fields to just
the metric and the dilaton. 

     In the gravity-scalar sector, the ansatz for the field
strengths can be rewritten in terms of the potentials, since we can
then write $\hat A_\3 = \ft1{4\sqrt2} g^{-3}\, s^{4/3}\, (3+2 c^2\,
\Delta^{-1}\, X^{-3})\, \ep_\3$, as may be seen from (\ref{fans}).
Substituting this and the other non-zero ans\"atze into the
ten-dimensional Lagrangian (\ref{romans1}) gives ${\cal L}_{10} =
\ft12 g^{-4}\, s^{1/3}\, c^3\, (R - \ft12(\del\phi)^2 + W)\,
\sqrt{-g_6}$, where
\bea
W &=& -\ft1{36} g^2 s^{-2}\, \Delta^{-2}\, X^{-12}\, 
 \Big(8s^2 +6s^4(1-27s^2) X^4 
+6s^2(4-45s^2+38s^4) X^8 \nn\\
&&\qquad\qquad\qquad\qquad\quad\quad 
 - c^2(1+118 s^2 -2s^4) X^{12} -72 s^2c^4 X^{16}\Big)  \,.
\eea
Although the $R$ and $(\del\phi)^2$ terms have a uniform
$\xi$-dependent prefactor, the term $W$, associated with the scalar
potential, does not.  Integration over the internal coordinate 
$\xi$ does not really make sense, since there is a divergence at
$\xi=0$.  One can make a suitable regularisation and thereby obtain
the scalar potential as given in (\ref{d6lag}), but this is
unsatisfactory since the result is scheme-dependent.\footnote{The
occurrence of the divergence is associated with the fact that the
metric in (\ref{fans}) has the warp factor $(\sin\xi)^{1/12}$, which
vanishes at $\xi=0$.  This singular behaviour is an inherent feature
of the massive type IIA theory, resulting from the scalar potential
$e^{\fft52\hat\phi}$, which has no stationary point.  (Since the
dilaton also diverges as $\xi$ aproaches zero, implying a passage to
the strong-coupling regime of the type I string theory, the effective
supergravity will in any case receive modifications.)  An analogous
calculation in the $S^4$ reduction of eleven-dimensional supergravity,
where no $\xi$-dependent warp factor arises, is free from any singular
behaviour \cite{lpd7gauge}.} Moreover, when the
higher-degree fields of the six-dimensional theory are included, it is
no longer possible to rewrite the ansatz for $\hat F_\4$ in
(\ref{fans}) as an ansatz for $\hat A_\3$.  We thus expect in this
case that one would not be able to obtain the six-dimensional
Lagrangian (\ref{d6lag}) by substituting the ans\"atze into the
ten-dimensional one.  It should be emphasised, however, that this is
not a drawback in the reduction procedure; rather, it just serves to
illustrate that Kaluza-Klein reduction is in general rather more
subtle than in the simple case of toroidal reduction.  The key point
is that given a consistent reduction, one has a way of embedding
solutions of the lower-dimensional equations of motion as solutions of
the higher-dimensional ones.

    An example of such a six-dimensional solution is an AdS black
hole, supported by a single component of the $SU(2)$ Yang-Mills
fields.  We find that the solution is given by
\bea
ds_6^2 &=& - H^{-3/2}\, f\, dt^2 + H^{1/2}\, (f^{-1}\, dr^2 + r^2\,
d\Omega^2_{4,k}) \,,\nn\\
\phi&=& \ft1{\sqrt2}\, \log H\,,\qquad A^{\sst 3}_\1 = \sqrt{2k} (1-H^{-1})
\,  \coth\beta\, dt\,,\nn\\
f &=&k -\fft{\mu}{r^3} + \ft29 g^2\, r^2\, H^2\,,\qquad H=1 + \fft{\mu\,
\sinh^2\beta}{k\, r^3}\,,\nn
\eea
where we have, for definiteness, chosen to use the $i=3$ component of
the Yang-Mills fields $A_\1^i$.  Another example is a supersymmetric
domain wall, supported by the scalar potential \cite{lpss}.  It is
straightforward to oxidise these solutions to ten dimensions, using
our ans\"atze (\ref{fans}).  If the parameter $\mu$ is set to zero in
the AdS black-hole solution, the six-dimensional metric becomes simply
AdS$_6$, and, as we remarked previously, the oxidation to $D=10$ gives
the near-horizon limit of the localised D4-D8 brane configuration (in
the case $k=0$).  When $\mu$ is instead non-zero, we expect that the
ten-dimensional interpretation will be that the D4-D8 brane system
will acquire a rotation, with angular frequency equal to the
black-hole charge, analogous to the cases discussed in
\cite{gubcve,ten}.   

    Another six-dimensional solution is the non-supersymmetric AdS$_6$
\cite{romans6}, corresponding to the second stationary point
$X=3^{-1/4}$ of the potential.  It is interesting to note that the
factor $\Delta$ appearing in the metric anstaz (\ref{fans}), which
takes the value $\Delta=1$ in the $X=1$ supersymmetric AdS$_6$
solution, now takes the form $\Delta=3^{-1/4}\, (1+2\sin^2\xi)$,
implying a distortion of the 4-sphere.

    To summarise, we have derived the gauged six-dimensional
supergravity by performing a consistent Kaluza-Klein reduction of
massive type IIA supergravity.  (For the sake of simplicity, we
concentrated on the full bosonic sector of the theories; the fermionic
sector will be addressed elsewhere.)  The metric ansatz describes a
warped product of the six-dimensional spacetime and a 4-sphere.  The
warp factor depends on the latitude coordinate of the 4-sphere, viewed
as a foliation of 3-spheres.  Since it vanishes on the equator, the
geometry of the internal space is really the upper hemisphere of the
4-sphere, with the equator as boundary.  (This is the region where
type I string theory becomes strongly coupled, and on the dual weakly
coupled heterotic string theory side a gauge enhancement takes place.)
We presented examples of six-dimensional solutions that can now be
re-interpreted as solutions of the massive IIA theory.  More
generally, our construction opens the door to the higher-dimensional
re-interpretation of any solution of the six-dimensional theory,
including, for example, non-abelian configurations.

\section*{Acknowlegement}

     We are grateful to T.A. Tran for useful discussions.


\begin{thebibliography}{99}

\bibitem{mald} J. Maldacena, {\sl The large $N$ limit of
superconformal field theories and supergravity},
Adv. Theor. Math. Phys. {\bf 2} (1998) 231, hep-th/9711200.

\bibitem{gkp} S.S. Gubser, I.R. Klebanov and A.M. Polyakov, {\sl Gauge
theory correlators from non-critical string theory}, Phys. Lett. {\bf
B428} (1998) 105, hep-th/9802109.

\bibitem{wit2} E. Witten, {\sl Anti-de Sitter space and holography},
Adv. Theor. Math. Phys. {\bf 2} (1998) 253, hep-th/9802150.

\bm{mald2} O. Aharony, S.S. Gubser, J. Maldacena, H. Ooguri and Y. Oz,
{\sl Large $N$ field theories, string theory and gravity}, hep-th/9905111.

\bm{romans6} L.J. Romans, {\sl The F(4) gauged supergravity in six
dimensions}, Nucl. Phys. {\bf B269} (1986) 691.

\bm{seiberg} N. Seiberg, {\sl Five dimensional SUSY field theories,
non-trivial fixed points and string dynamics}, Phys. Lett. {\bf B388}
(1996) 753, hep-th/9608111.

\bm{intrill} K. Intriligator, D.R. Morrison and N. Seiberg,
{Five-dimensional supersymmetric gauge theories and degenerations
of Calabi-Yau spaces}, Nucl. Phys. {\bf B497} (1997) 56, hep-th/9702198.

\bm{ferrara} S. Ferrara, A. Kehagias, H. Partouche and A. Zaffaroni,
{\sl AdS$_6$ interpretation of 5-D superconformal field theories},
Phys. Lett. {\bf B431} (1998) 57, hep-th/9804006. 

\bm{romanstype2a} L.J. Romans, {\sl Massive $N=2a$ supergravity in ten
dimensions}, Phys. Lett. {\bf B169} (1986) 374.

\bm{oz} A. Brandenhuber and Y. Oz, {\sl The D4-D8 brane system and
five dimensional fixed points}, hep-th/9905148.

\bm{youm} D. Youm, {\sl Localised intersecting BPS branes},
hep-th/9902208.

\bm{branetension} I.V. Lavrinenko, H. L\"u, C.N. Pope and K.S. Stelle,
{\sl Superdualities, brane tensions and massive IIA/IIB duality},
hep-th/9903057, to appear in Nucl. Phys. {\bf B}.

\bm{lpss} H. L\"u, C.N. Pope, E. Sezgin and K.S. Stelle, {\sl
Dilatonic $p$-brane solitons}, Phys. Lett. {\bf B371} (1996) 46, 
hep-th/9511203. 

\bibitem{gubcve} M. Cveti\v{c} and S.S. Gubser, {\sl Phases of
R charged black holes, spinning branes and strongly coupled gauge
theories}, hep-th/9902195.

\bm{ten} M. Cveti\v{c}, M.J. Duff, P. Hoxha, J.T. Liu, H. L\"u,
J.X. Lu, R. Martinez-Acosta, C.N. Pope, H. Sati and T.A. Tran, {\sl
Embedding AdS black holes in ten and eleven dimensions},
hep-th/9903214, to appear in Nucl. Phys. {\bf B}.

\bm{lpd7gauge} H. L\"u and C.N. Pope, {\sl Exact embedding of $N=1$,
$D=7$ gauged supergravity in $D=11$}, hep-th/9906168.



\end{thebibliography}
\end{document}